# Vibrational approaches for symmetrization of laminar viscous incompressible fluid flow in a flat diffuser

**A. I. Fedyushkin[1], A.A. Puntus[2]**

1. Ishlinsky Institute for Problems in Mechanics of the Russian Academy of Sciences, Moscow, Russia

2. Moscow Aviation Institute (National Research University), Moscow, Russia

E-mails: fai@ipmnet.ru, artpuntus@yandex.ru

**Abstract**. The article shows two methods of symmetrization of the asymmetric flow of a viscous incompressible fluid in a flat diffuser. The first is the effect of a weak periodic vibration on the input stream. The second is by means of a vibration effect with a small amplitude on the flow velocity from the walls of the diffuser. The results are obtained for a viscous incompressible fluid by modeling based on the numerical solution of the Navier-Stokes equations.

## 1. Introduction

The problem of the flow of a viscous incompressible fluid in a flat diffuser for small Reynolds numbers and the symmetric case was solved independently by Jeffrey [1] and Hamel [2] more than a hundred years ago. It is known that the flow of a viscous incompressible fluid in a flat diffuser at low Reynolds numbers is symmetric, but as the Reynolds number increases above the critical number Re*, the flow loses symmetry, remaining stationary, and when the second critical Reynolds number is exceeded, the asymmetric flow becomes oscillatory [3], and with a further increase of the Reynolds number, the flow goes into the turbulent mode.

The study of nonlinear modes of laminar fluid flows in a diffuser, for example, such as asymmetry and intermittency of the flow, is of great fundamental and applied importance. However, the asymmetry problem of laminar stationary flows in the diffuser has not been sufficiently studied, and few papers have been devoted to this problem compared to a lot of articles on turbulent flows. A review of scientific works on the solution of the Jeffrey–Hamel problem and a generalization based on group analysis are given in the paper [4]. The results paper [4] indicates existence a non-uniqueness of the stationary solutions of the Jeffrey–Hamel problem, i.e., the possibility an appearance of stationary asymmetry of the fluid flow in the diffuser. The authors of the papers [5, 6] found generalizations of the solution of the Jeffrey-Hamel problem, presented one-, two- and three-mode bifurcation solutions indicating the presence of asymmetric stationary flows for certain ranges of Reynolds numbers and diffuser opening angles. In the papers [3, 7, 8] laminar symmetric and asymmetric stationary and transient flow modes in a flat diffuser with a small opening angle were studied based on the numerical solution of the Navier-Stokes equations for a viscous incompressible fluid. The range of Reynolds numbers for the existence of these modes of fluid flow in the diffuser was indicated. The changes in the structure of flows from stationary-symmetric to stationary-asymmetric and to non-stationary in the diffuser and confuser depending on the Reynolds number for Newtonian and non-Newtonian liquids with the Ostwald-de Waele power law of viscosity were shown in the article [9]. In this article also the results of laminar flows of viscous fluid in a flat diffuser and confuser with symmetrical and asymmetric velocity profiles at the inlet boundary were presented.

In addition to fundamental importance, studies of vibration effects on the flow of viscous incompressible fluid in flat channels are also of applied importance, for example, in biomedical and technological applications. Many interesting features were found in the oscillating flows of viscous fluid in the channels. For example, in the viscous fluid flowing through pipe there is Richardson's "annular effect" [10] when on inlet fluid flow influence periodic fluctuations. The effect is that when vibrations with relatively high frequencies are superimposed on the flow, a maximum of the average longitudinal velocity occurs in a narrow layer of fluid near wall of tube. At the same time, in the rest of the pipe, the fluid oscillates (as a solid) in accordance with the fluctuations of the average cross-section velocity. If the vibrations are removed, this effect will disappear. The results of the paper [11] showed that acoustic vibrations change gas flow velocity and temperature in the zone of the boundary layers of a convective jet, where two temperature maxima appear. In a paper [12] the effect of the opening angle and the extension of the diffuser on the asymmetry of the flow in a flat diffuser is numerically shown, and it is told that the imposition of periodic vibrations on the input flow can symmetrize the flow, but this needs a more detailed study.

Our article is a continuation of the works [3, 7, 8, 9] and presents the results concerning the method of extending the range of Reynolds numbers for symmetric flows by imposing a weak vibration effect on the input velocity of fluid flow. In the paper, a technique of symmetrization of an asymmetric flow of a viscous incompressible fluid in a flat diffuser using different periodic vibration effect on the velocity of the input flow $V_{in}$ is shown. The results of modeling the flow of a viscous incompressible fluid based on the numerical solution of the Navier-Stokes equations for various Reynolds numbers (in the range up to $10^3$) and various periodic vibration effects on the velocity $V_{in}$ at the inlet to the diffuser ($V = V_{in} + A\sin(2\pi f t)$), here A is the amplitude, f is the frequency) are presented.

## 2  The problem statement

The laminar flow of a viscous incompressible fluid driven through a channel bounded by two flat walls inclined towards each other at a small angle β is considered. In this paper we consider flat diffuser bounded by two arcs ("input" and "output" boundary) with the one center (Fig. 1a). The entry flow in the classical problem of Jeffrey - Hamel about symmetric flow in a flat diffuser was point's source and arc was for output flow. The aim of our numerical simulation is to determine the effect of vibration on the asymmetric flow in the diffuser in order to symmetrize it.

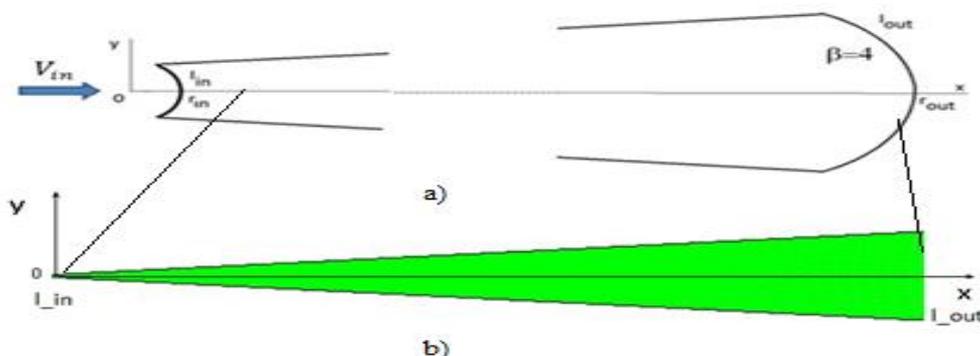

**Figure 1.** Scheme of the computational domain for a flat diffuser: a) enlarged parts of the computational domain near the inlet (on the left) and outlet (on the right) of the diffuser; b) the computational domain with an example of grid ($\beta = 4°, L = 0.495$ m)

The geometry of the mathematical model was chosen the same as the geometry of the diffuser of Jeffrey Hamel's problem. This was chosen in order to be able to compare our results with the results of well-known works [1, 2, 4-6]. Geometric model of the diffuser is as follows: opening angle is $\beta = 4°$, the input boundary has the form of an arc $l_{in}$ ($r_{in}$=0.005 m) where $r$ is calculated by formula $r^2=x^2+y^2$. The shape of the output boundary also has form of an arc $l_{out}$ ($r_{out}$=0.5 m) what was shown in Fig. 1a. The length of the diffuser $L$ is equal to 0.495 m, and is calculated by formula $L=|r_{out} - r_{in}|$. The simulation of the problem is carried out on the basis of the numerical solving of the two-dimensional Navier – Stokes equations for an incompressible viscous fluid.

$$\frac{\partial V}{\partial t} + (V\nabla)V = -\frac{\nabla P}{\rho} + \nu\Delta V, \quad \text{div} V = 0$$

where $V$ is the velocity vector; $P$ is the pressure; $\rho$ is the density; $t$ is the time.

As the boundary conditions we take: at the inlet of the diffuser velocity is a constant $V_{in}$ (Reynolds number $\text{Re}_{in}$) or $V = V_{in} + A\sin(2\pi f t)$, at the arc of output the pressure $P = 0$ is assumed, at the upper and lower walls for velocity no slip conditions $V = 0$ are assumed. The initial conditions are $t = t_0 = 0$, $V(t_0) = 0$, $P = 0$. The velocity scale is chosen by the velocity $V_{in}$ and the Reynolds numbers are defined as $\text{Re} = \text{Re}_{in} = V_{in} l_{in} / \nu$, $\text{Re}_{vibr} = A l_{in} / \nu$, $y\_{dimless}=y/r \sin(\beta/2)$, $Vx\_{dimless}=Vx/Vx_{in}$, $Vy\_{dimless}=Vy/Vx_{in}$.

For the numerical solution of Navier - Stokes equations were used the finite-difference and the control volume methods. Numerical calculations were carried out with using the method control volumes with schemes second and third order accuracy on space and second order on time's approximation. The validations of numerical calculations were confirmed on solutions of convection problems by comparing the results obtained using various numerical methods and comparing them with experimental data. The results of validations of numerical models were published in article (Fedyushkin and Puntus 2018). Test calculations were carried out on sequence of the meshes with decreasing grid step. The grids were selected, for which the results did not differ as number of grid nodes increased. The results for quasi-stationary (with stationary for average values) flow modes were obtained on grids with a margin (the values of the grid Reynolds number and the Courant number were no more than unity). The number of nodes was at least $10^5$. The meshes were rectangular and orthogonal near solid walls. We used uneven grids with decreasing mesh spacing at the input of the diffuser and near solid boundaries (where ten nodes of the mesh there were within the boundary layers at least). For a given input velocity, calculations were performed from zero initial velocities in the computational domain until an establishment of a steady-state (or quasi-steady-state) flow mode. The analysis of the numerical solutions was carried out for the steady-state (or quasi-steady-state) of flow regimes. In Fig. 1 the computational domain is shown. Geometric details near the input and output walls with radius of curvature $r_{in}$ and $r_{out}$ are shown in Fig. 1a. The computational domain and one of the grids used in these calculations are presented in Fig. 1b.

## 3 The numerical results

The simulation results are presented in Fig. 2-7 in the form of isolines and velocity profiles.

**3.1 The fluid flows in the diffuser without vibrations ($V_{in} \neq 0$, $A = 0$)**

The results for the cases of symmetrical (Re = 249 and Re = 268) and asymmetric fluid flows (Re=279 and Re=499) in the diffuser without superimposed periodic effects are presented in Fig. 2 and Fig. 3 (Fedyushkin 2016). The results for a case of symmetric flows and the value of critical number Re * of transition to asymmetric fluid flow for cases without influence of vibrations well coincide with results of works [5,6].

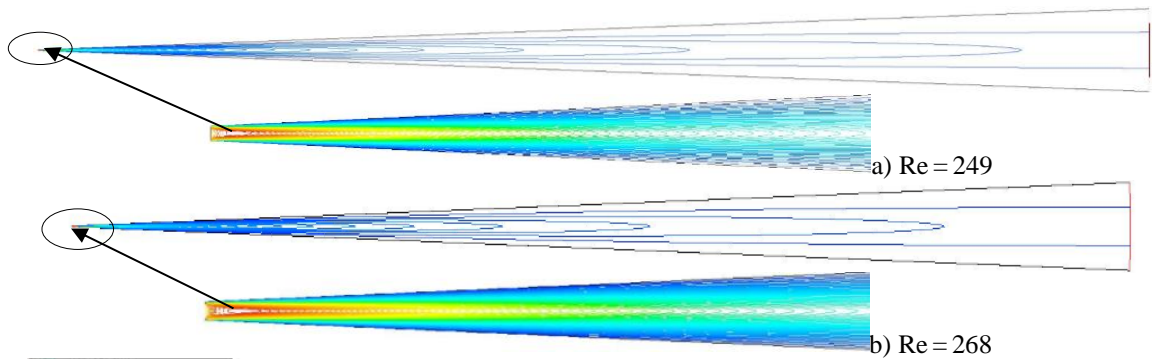

a) Re = 249

b) Re = 268

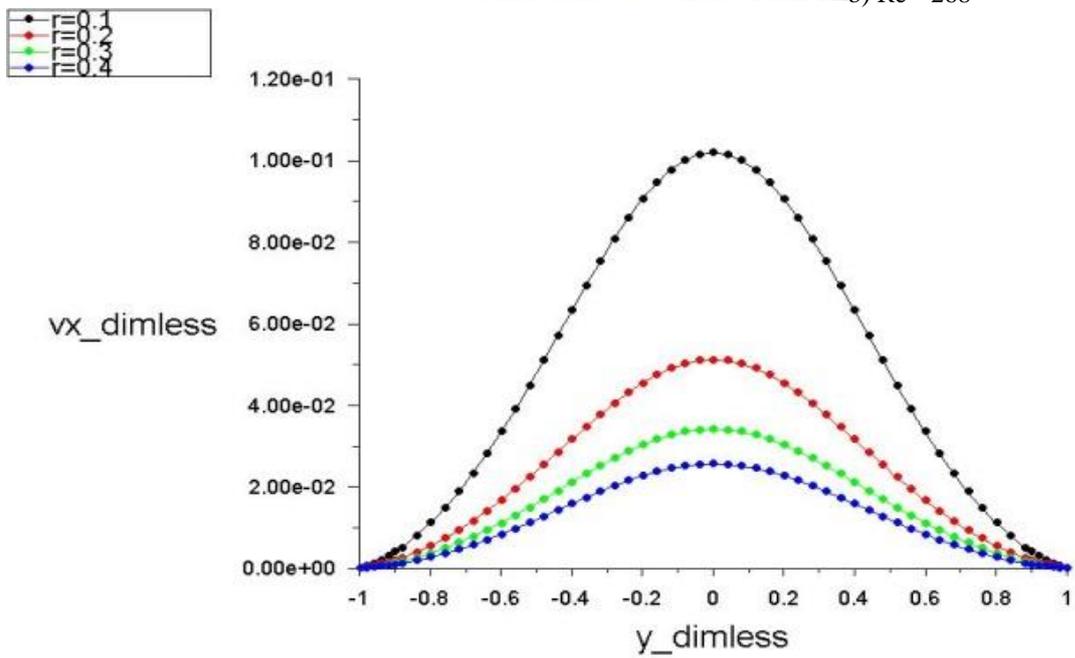

c)

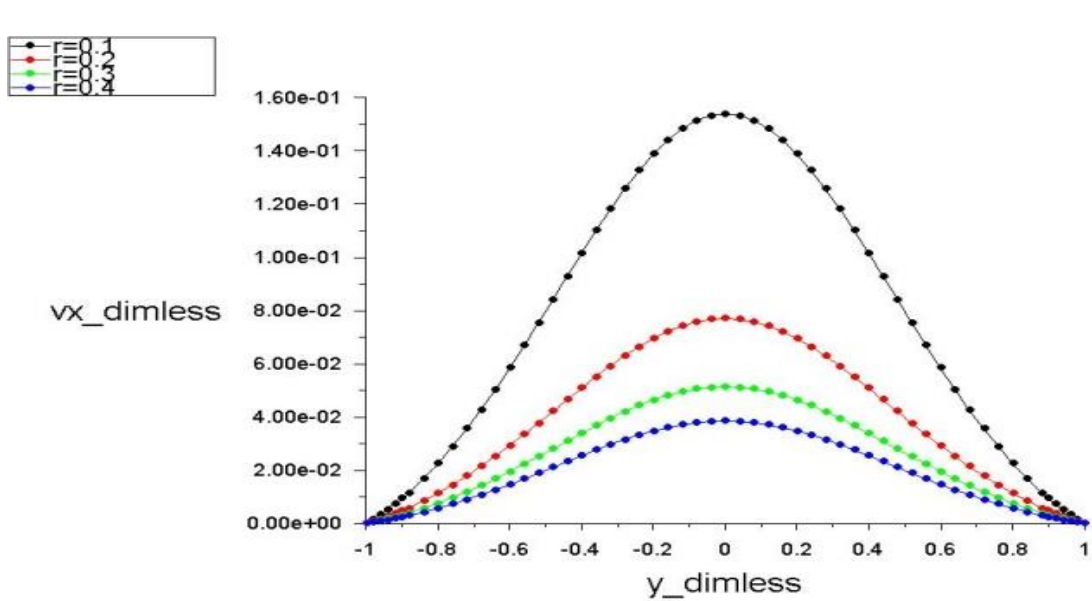

d)

**Figure 2.** The isolines of the horizontal component of the velocity vector for the case of symmetrical fluid flows: a) Re = 249; b) Re = 268 and vertical profiles of the velocity $V_x$

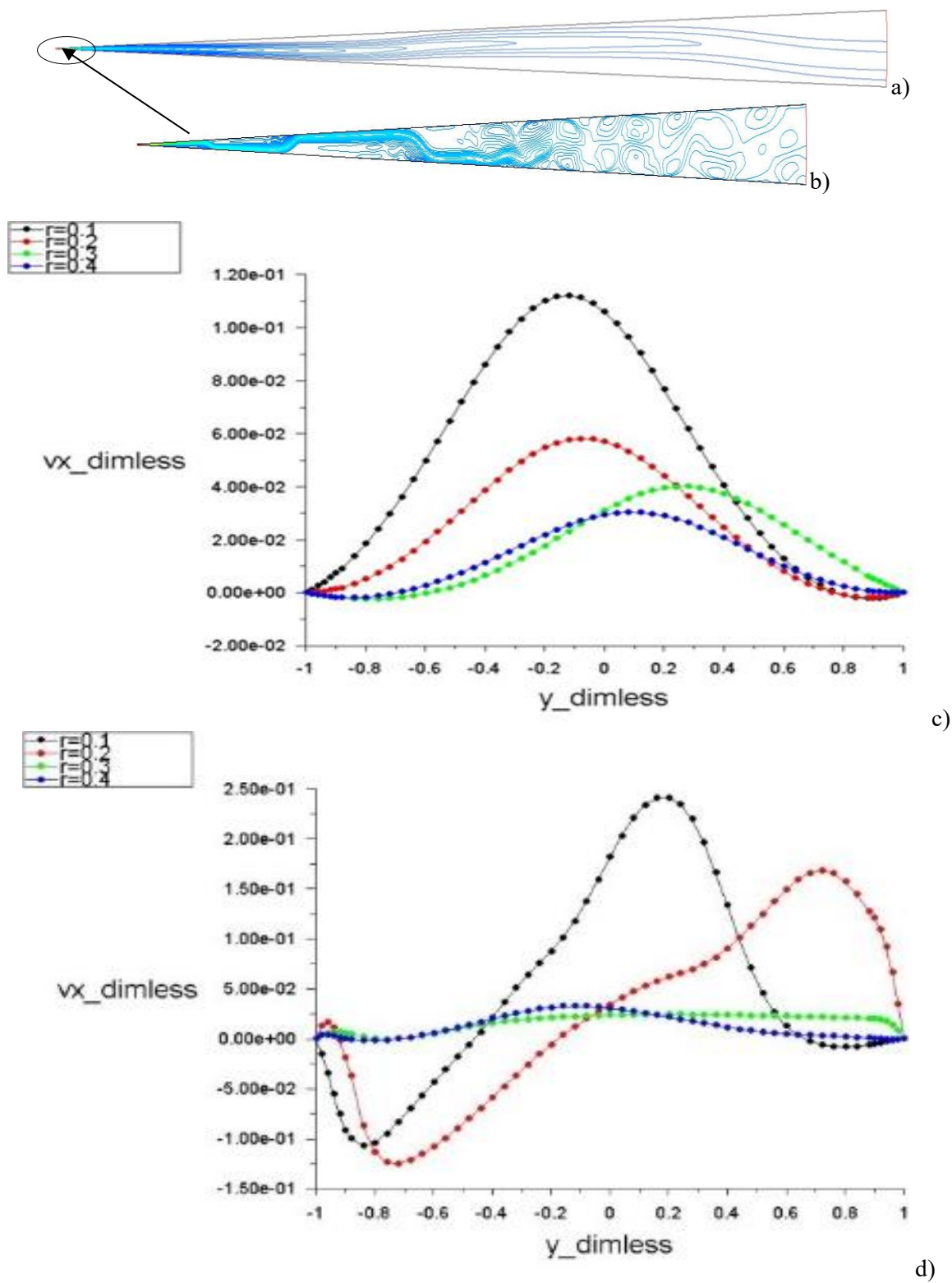

**Figure 3.** The isolines and the profiles in vertical cross-sections of horizontal component $V_x$ of velocity vector for the case of asymmetrical fluid flows: a), c) Re = 279; b), d) Re = 499

## 3.2 Only vibrational fluid flow in the diffuser ($V_{in} = 0$, $A = 1$ m/s, $f = 10$ Hz)

Figure 4 shows the patterns of the fluid flow for the case only with a periodic change in velocity at the entrance to the diffuser ($V_{in} = 0$, $A = 1$ m/s, $f = 10$ Hz, $Re_{vibr}$=349). On the average velocity profiles, you can notice the velocity maxima near the walls – this is the Richardson's "annular effect" [10].

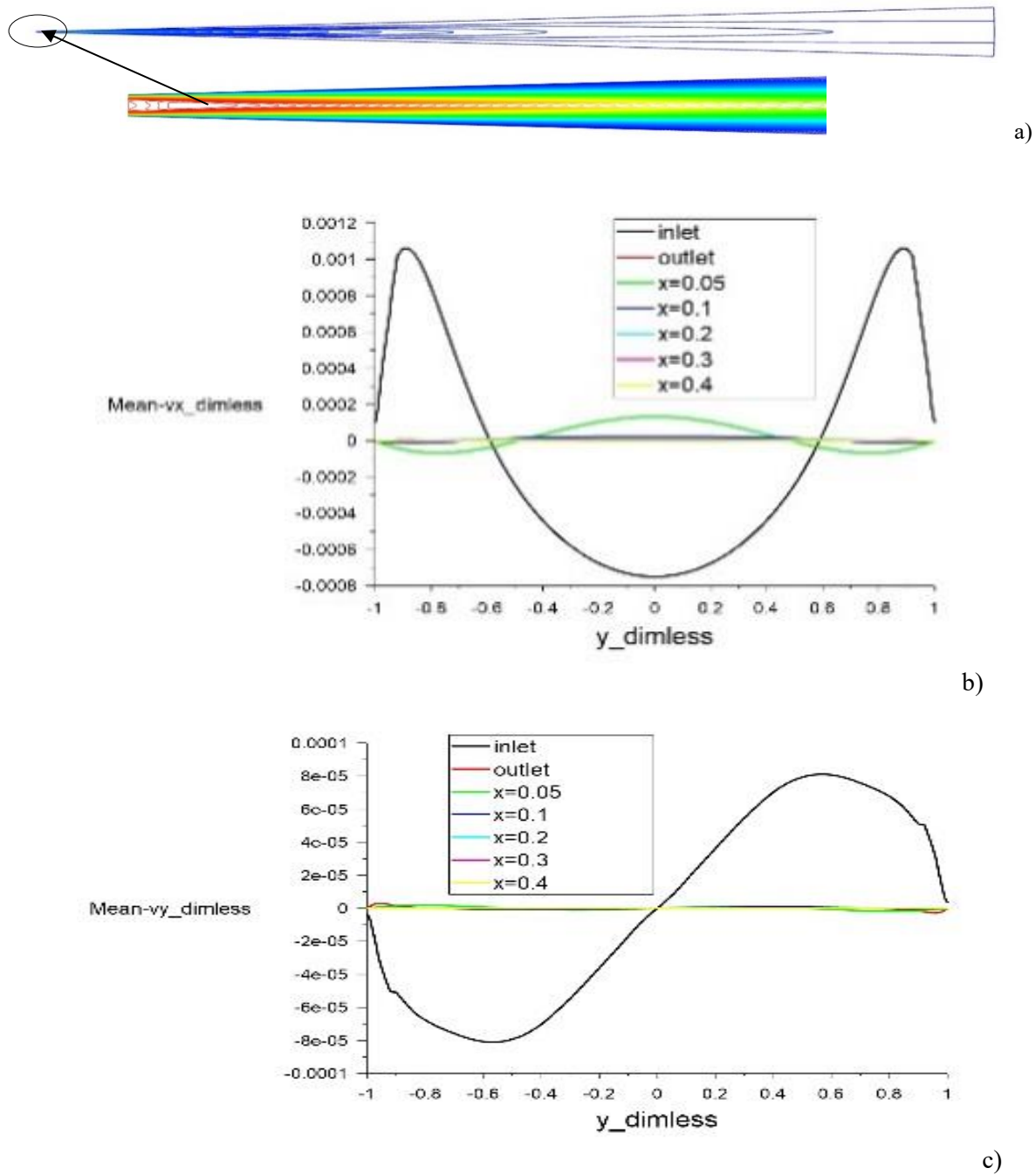

**Figure 4.** The isolines of the averaged longitudinal component of the velocity mean_$V_x$, (below are the isolines of the mean_$V_x$ velocity near the entrance to the diffuser) (a), the profiles of the longitudinal component mean_$V_x$ of the velocity (b) and the averaged transverse component mean_$V_y$ of velocity in cross-sections (c) for case $V_{in} = 0$, $A = 1$ m/s, $f = 10$ Hz

## 3.3 The fluid flow in the diffuser with a vibrational effect ($V_{in} \neq 0$, $A \neq 0$)

The effect of a periodic disturbance on the mainstream (Re=279) with a frequency f = 10 Hz for two amplitudes: A=0.1 m/s (Re$_{vibr}$=2.4) and A=10 m/s (Re$_{vibr}$=240) are presented in Fig. 5 and Fig. 6. Comparison of the results in Fig.3 and Fig. 5 shows that the influence of vibrations, even at amplitudes less than 1% of the velocity $V_{in}$ can lead to symmetrization of the fluid flow in the diffuser.

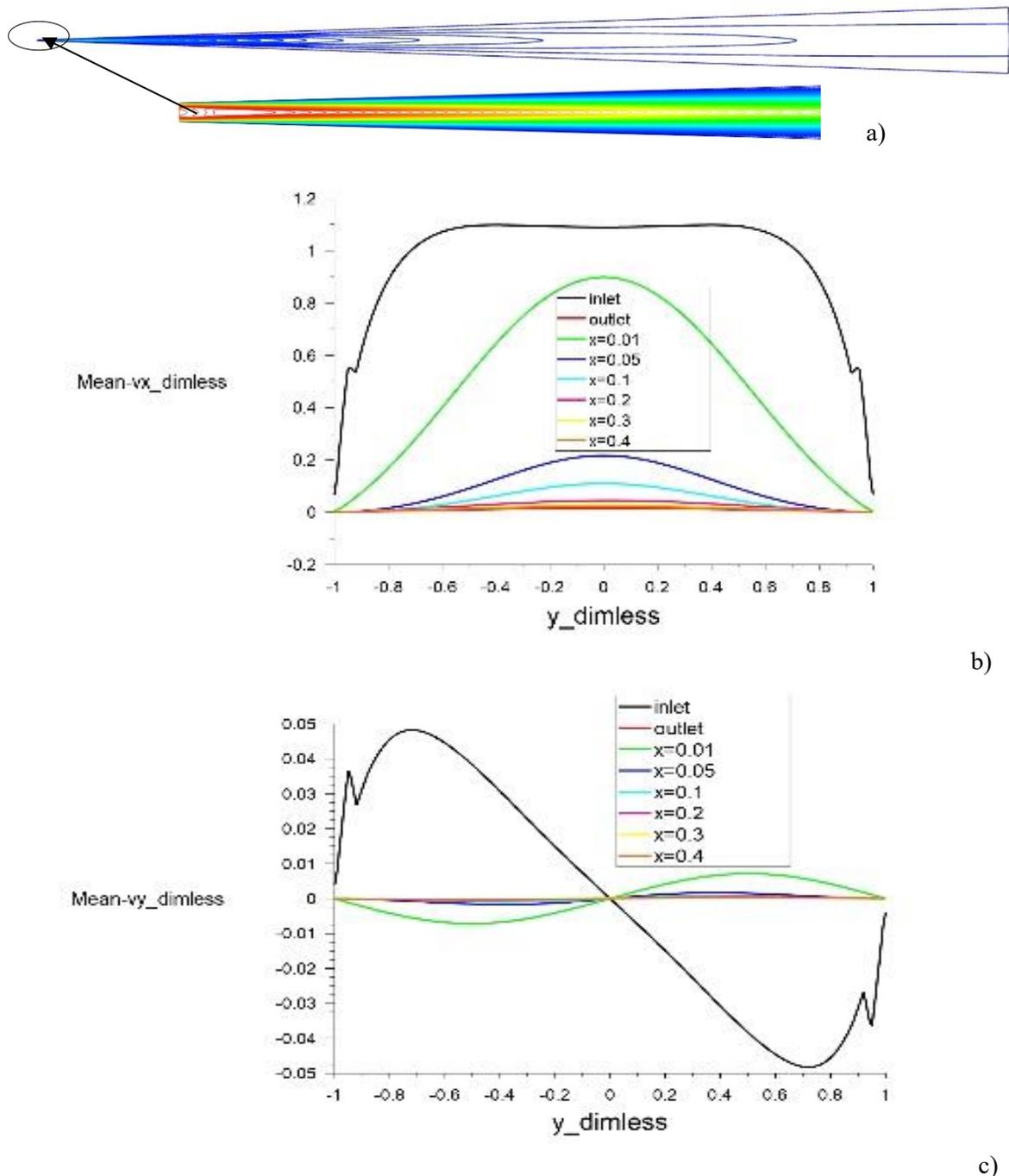

**Figure. 5.** The isolines of the averaged longitudinal component of the velocity mean_V$_x$, (below are the isolines of the mean_V$_x$ velocity near the entrance to the diffuser) (a), the profiles of the longitudinal component of the mean_V$_x$ velocity (b) and the transverse component of the mean_V$_y$ velocity in cross-sections (c) for case $V_{in} = 11.7\,\text{m/s}, A = 0.1\,\text{m/s}, f = 10\,\text{Hz}$ (Re=279, Re$_{vibr}$=2.4)

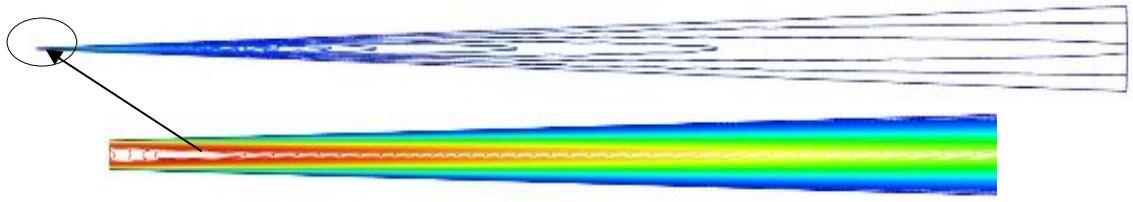

a)

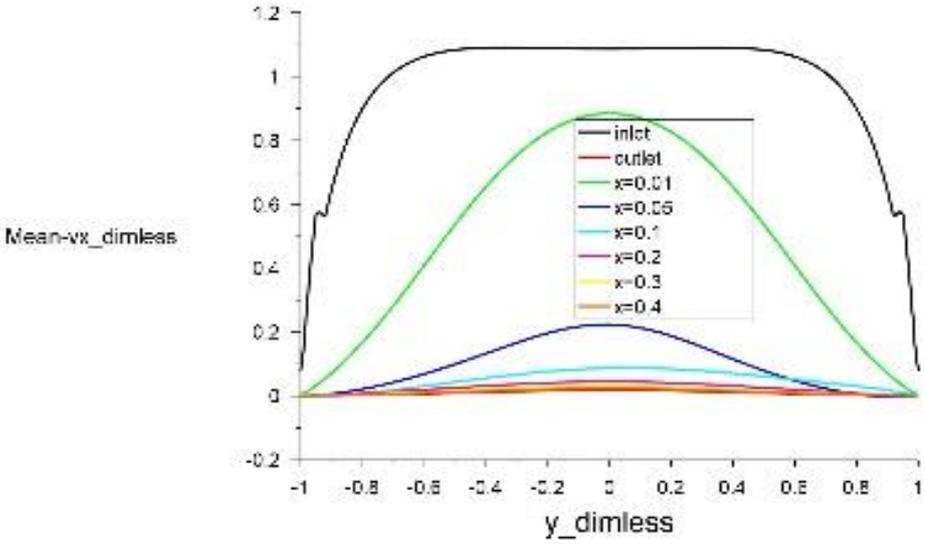

b)

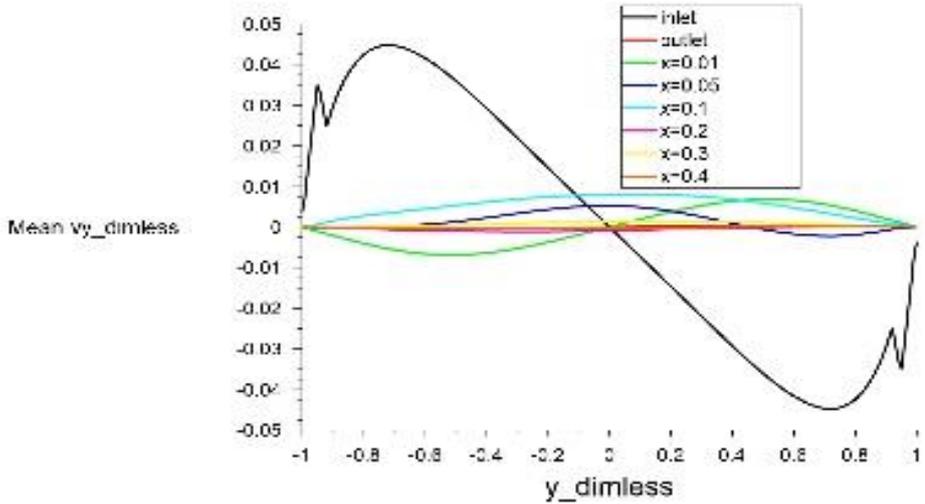

c)

**Figure 6.** The isolines of the averaged longitudinal component of the velocity mean_$V_x$, (the isolines of the mean_$V_x$ velocity near the entrance to the diffuser are presented in lower figure) (a), the profiles of the longitudinal component of the mean_$V_x$ velocity (b) and the transverse component of the mean_$V_y$ velocity in cross-sections (c). $V_{in} = 11.7 \, m/s, A = 10 \, m/s, f = 10 Hz$, (Re=279, Re$_{vibr}$=240).

The results in Figures 5 and 6 are almost identical in terms of flow symmetry, which shows that the frequency of vibrational action has little effect on the symmetrization of the average flow velocity

An example second approach of symmetrization of the fluid flow velocity in a flat diffuser by vibration action along normal to the walls of the diffuser according to the harmonic law $V_n = A \sin(2\pi f)$ with a small amplitude A and a frequency f is shown in Fig. 7. In Figure 7 mean_Vx – is the time-average velocity profiles (Fig 7a) and rmse_Vx – is the deviation profiles from the dimensionless average velocity (Fig 7b) for Re=279, A=0.001m/s, f=10 Hz ($Re_{vibr}$=0.02) are shown.

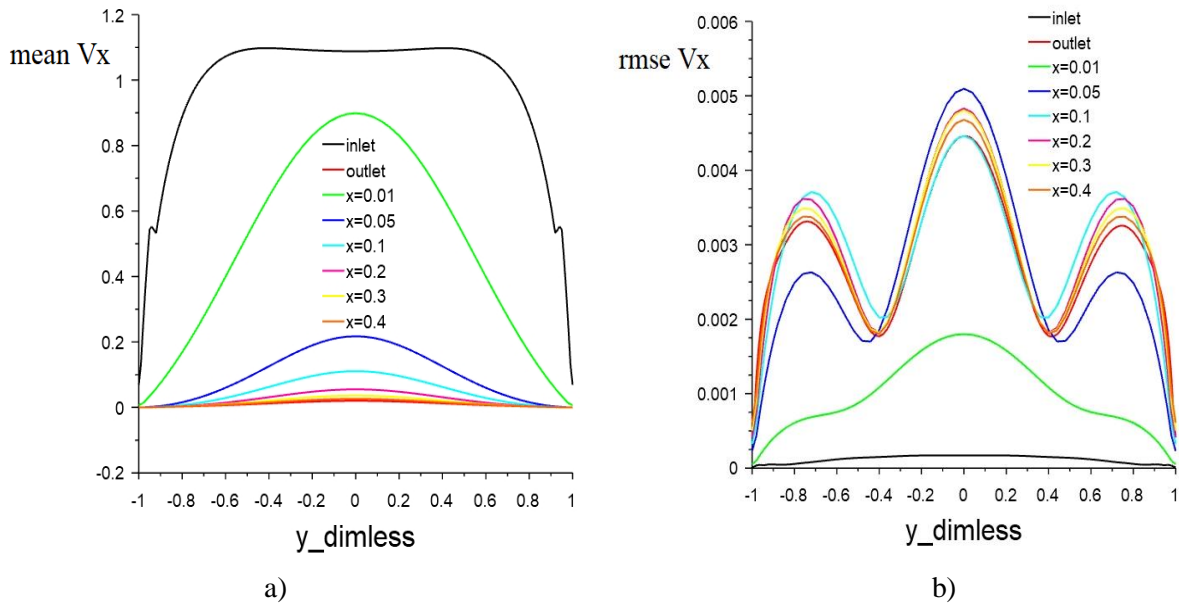

a)  b)

**Figure 7**. The profiles of time average velocity (mean_Vx) (a) and deviation profiles from average velocity (rmse_Vx) (b) for fluid flow in a flat diffuser with vibration action from the walls of the diffuser for Re=279, A=0.001m/s, f=10Hz ($Re_{vibr}$=0.02)..

## 4 Conclusion

The results of numerical simulation have shown two ways of symmetrization of asymmetric laminar flows of viscous incompressible fluid in a flat diffuser: the first - due to a weak periodic effect on the flow velocity at the entrance to the diffuser and the second - due to vibration action from the walls of the diffuser. It is shown that the impact of vibration, even at amplitudes less than 1% of the velocity $V_{in}$, can lead to the symmetrization of the fluid flow in the diffuser. Richardson's "ring effect" (that is, the effect of the influence of harmonic oscillations of the input flow on the shape of the fluid flow velocity profile in a cylindrical pipe) was demonstrated for a diffuser.

**Acknowledgements**

*This work was supported by the Russian Science Foundation grant 24-29-00101.*